\documentclass[a4paper,traditabstract]{aa}  
\usepackage{graphicx,amsmath,txfonts,natbib,astro,url,amssymb,amsmath}
\usepackage[version=3]{mhchem}
\usepackage[normalem]{ulem}
\begin{document}
\def\iras{IRAS16293-2422}
\def\ncrit{\ensuremath{n_{\rm crit}}}
\def\tc {\ensuremath{T_C}}
\def\ntot{\ensuremath{N_{\rm tot}}}
\def\aul{\ensuremath{A_{ul}}}
\def\gu{\ensuremath{g_{u}}}
\def\tauk{\ensuremath{\tau_{ul}}}
\def\tant{\ensuremath{{T_{\rm ant}}}}
\def\jnu{\ensuremath{J_\nu}}
\def\tcmb{\ensuremath{T_{\rm CMB}}}
\def\jnuex{\ensuremath{{\tilde{T}_{\rm ex}}}}
\def\jnucmb{\ensuremath{{\tilde{T}_{\rm cmb}}}}
\def\osu{\texttt{osu.09.2008}}
\def\tfm#1{\tablefootmark{#1}}
\def\tft#1#2{{\tablefoottext{#1}{#2}}}
\def\tfoot#1{{\tablefoot{\scriptsize #1}}}
\def\ratio{\ce{NH}:\ce{NH2}:\ce{NH3}}
\def\corr#1{\textbf{#1}}
\def\reac#1{(\ref{r:#1})}
\def\ab#1{\ensuremath{{[\ce{#1}]}}}
\def\xref{\textbf{REFERENCES}}
\def\ri{{\ce{NH}:\ce{NH2}}}
\def\rii{{\ce{NH3}:\ce{NH2}}}
\def\opr{\ensuremath{\rm O/P}}
\def\ohh{\ce{o-H2}}
\def\phh{\ce{p-H2}}
\def\nhh{\ce{n-H2}}
\title{Nitrogen hydrides and the H$_2$ ortho-to-para ratio in dark clouds}

\titlerunning{}

\author{%
V. Dislaire     \inst{1}  	\and
P.~Hily-Blant	\inst{1}	\and
A.~Faure        \inst{1}	\and
S.~Maret	\inst{1}	\and
A.~Bacmann	\inst{1}        \and
G. Pineau des For\^ets \inst{2}}

\institute{$^1$ Universit\'e Joseph Fourier and CNRS, Institut de
  Plan\'etologie et d'Astrophysique, Grenoble, France\\$^2$ Universit\'e
  Paris-Orsay and CNRS, Institut d'Astrophysique Spatiale, Orsay,
  France }

\abstract{Nitrogen bearing species are common tracers of the physical
  conditions in a wide variety of objects, and most remarkably in dark
  clouds. The reservoir of gaseous nitrogen is expected to be atomic
  or molecular, but none of the two species are observable in the dark
  gas. Their abundances therefore derive indirectly from those of
  N-bearing species through chemical modelling. The recent years have
  accumulated data which stress our incomplete understanding of the
  nitrogen chemistry in dark cloud conditions.  To tackle this problem
  of the nitrogen chemistry in cold gas, we have revised the formation
  of nitrogen hydrides, which is initiated by the key reaction \ce{N+
    + H2 -> NH+ + H}. We propose a new rate for this reaction which
  depends on the ortho-to-para ratio of \hh. This new rate allows to
  reproduce the abundance ratios of the three nitrogen hydrides, NH,
  \ce{NH2}, and \ce{NH3}, observed towards \iras, provided that the
  channel leading to NH from the dissociative recombination of
  \ce{N2H+} is not closed at low temperature.  The ortho-to-para ratio
  of \hh\ is constrained to $\opr=\dix{-3}$ by the abundance ratio
  \ri, which provides a new method to measure \opr. This work stresses
  the need for reaction rates at the low temperatures of dark clouds,
  and for branching ratios of critical dissociative recombination
  reactions.}



\maketitle


\section{Introduction}

Chemistry in an astrophysical context is not only a matter in itself
but also provides invaluable tools to determine physical conditions
such as volume density and kinetic temperature \citep{bergin2007}.
Astrochemistry relies on chemical networks which depend on the type of
environment (\eg\ diffuse \vs\ dense gas). The kinetic rates of the
reactions are, in the best cases, based on thermodynamical and quantum
mechanical calculations and experiments, which make the setup of such
chemical networks an extremely demanding process. Various such
networks of reactions are publicly available (KIDA, UMIST,
OSU\footnote{\url{http://kida.obs.u-bordeaux1.fr},
  \url{http://www.udfa.net/},
  \url{http://www.physics.ohio-state.edu/~eric}}, Flower \& Pineau des
For\^ets\footnote{\url{http://massey.dur.ac.uk/drf/protostellar/}}),
with various degrees of complexity and/or completeness regarding
specific aspects of the chemistry (\eg\ deuteration, cations...).

The nitrogen element is among the 5th or 6th most abundant in the
Solar neighbourhood, after H, He, C, O, and probably Ne
\citep{asplund2009, nieva2011}. In the cold neutral medium, nitrogen
is expected to be predominantly atomic or molecular, but direct
observations of N or \ce{N2} are not possible. The amount of gaseous
nitrogen thus relies on the abundances of N-bearing molecules \via\
chemical models. Nitrogen bearing molecules are observed in a wide
variety of physical conditions. Molecules such as CN, HCN, and HNC,
with large permanent dipole moments, are detected towards diffuse
clouds \citep{liszt2001}, dense cores \citep{tafalla2004},
protoplanetary disks \citep{kastner2008b} and high-z galaxies
\citep{guelin2007}. N-bearing species, such as \ce{N2H+} (and its
deuterated isotopologues), are also efficient tracers of the dense and
cold gas where CO has already frozen-out \citep{crapsi2007,
  hilyblant2008cn, hilyblant2010n}. The cyanide radical CN is also a
precious molecule which serves as a tracer of the magnetic fields
through Zeeman splitting \citep{crutcher2010}. Understanding the
chemistry of nitrogen is thus crucial in many astrophysical areas.

Ammonia and \ce{N2H+} are daughter molecules of \ce{N2}, which forms
from atomic N through neutral-neutral reactions mediated by CN and NO
\citep{pineau1990}. The formation of CN and the related HCN and HNC
molecules, and in particular their abundance ratio CN:HCN, are however
not fully understood \citep{hilyblant2010n}.  The formation of ammonia
in dark and dense gas is thought to take place in the gas phase. Once
\ce{N2} exists in the gas phase, it reacts with \ce{He+} ions formed
by cosmic-ray ionization, to form \ce{N+} which by subsequent hydrogen
abstractions lead to \ce{NH4+} and finally \ce{NH3} by dissociative
recombination. \cite{lebourlot1991} (LB91 in the remainder of the
paper) was the first to investigate the influence of the \ce{H2}
ortho-to-para ratio (noted \opr) on the formation rate of
\ce{NH3}. LB91 considered the conversion between \phh\ and \ohh\
through proton exchange in the gas phase. He concluded that when $\opr
\ge \dix{-3}$, the formation of ammonia no longer depends on the value
of \opr.  However, a better test of the nitrogen chemistry is provided
by the simultaneous observations of all three nitrogen hydrides, NH,
\ce{NH2}, and \ce{NH3} \citep[\eg,][]{persson2010}. The Herschel/HIFI
instrument has opened the THz window which contains the fundamental
rotational transitions of light molecules with large electric dipole
moments such as hydrides. \citet[][Paper~I in what
follows]{hilyblant2010nh} have detected NH, \ce{NH2}, and \ce{NH3} in
the cold envelope of the Class~0 protostar \iras. The hyperfine
structures of the $N=1-0$ transitions of NH and \ce{NH2} are seen for
the first time, in absorption, and allow a precise determination of
their excitation temperature (\texc) and line centre opacity. Four
lines of ammonia are also detected in absorption, plus the fundamental
rotational transition at 572~GHz which evidences a self-absorbed line
profile, likely tracing both warm and cold gas. For all three
molecules, excitation temperatures were found in the range 8--10~K
(Paper~I). The derived column density ratios are $\ratio =
5:1:300$. Whilst abundances are generally delicate to derive because
the column density of \ce{H2} is uncertain, the column density ratios
above provide a stringent test for the first steps of the nitrogen
chemistry and for the ammonia synthesis in particular. Indeed, the
current nitrogen networks available all produce more \ce{NH2} than
\ce{NH} under dark cloud conditions (Paper~I), at odds with the
observations.

In this Letter, we investigate the NH:\ce{NH2} problem in dark gas, by
revisiting the kinetic rate for the reaction
\begin{equation}
  \ce{N+ + H2 -> NH+ + H}
  \label{r:nhp}
\end{equation}
We derive separate rates for reaction with p- and \ohh\ for which the
nuclear spin $I=0$ and $1$ respectively. In Section 2 we describe the
new rates which are compared to experimental measurements and derive
the abundances of nitrogen hydrides in typical dark cloud
conditions. The results are discussed in Section~3 and we propose
conclusion remarks in Section~4.

\section{Chemical modelling}

\begin{figure}
  \centering
  \includegraphics[width=.9\hsize,angle=-90]{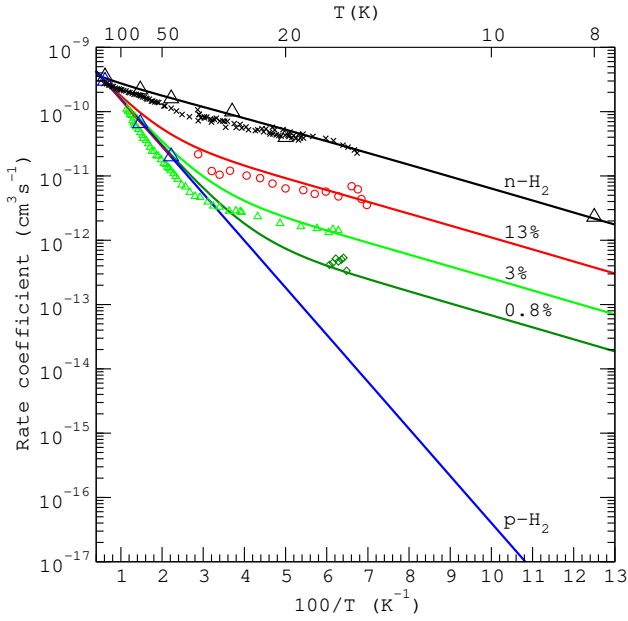}
  \caption{{Comparison of the rate coefficient as given by
      Eq.~\ref{eq:k1} to experimental data for the reaction of \ce{N+}
      with \nhh\ and \phh\ with different \ohh\ admixtures: 13\%, 3\%,
      0.8\%, and 0\% (blue line). The small symbols are taken from
      \cite{gerlich1993} while the large triangles are CRESU results
      from \cite{marquette1988}. The solid lines correspond to
      Eq.~\ref{eq:k1} of the present paper.}}
  \label{fig:exp}
\end{figure}

\begin{figure}
  \centering
  \includegraphics[width=.9\hsize]{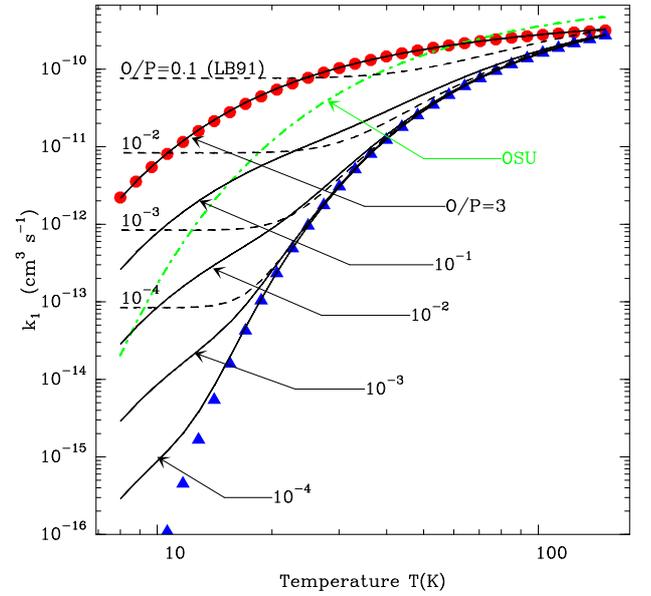}
  \caption{{Comparison of the rate coefficient for reaction \reac{nhp}
      as given by Eq.~\ref{eq:k1} to the rates from LB91 (dashed) and
      the OSU database (dot-dashed). The rates are computed for
      temperatures in the range 8--150~K and $\opr=\dix{-4}$,
      \dix{-3}, 0.01, 0.1, and 3 (black lines). The symbols show the
      fits from \cite{marquette1988} for normal \hh\ (red circles,
      \opr=3) and p-\hh\ (blue triangles).}}
  \label{fig:k1}
\end{figure}

\subsection{Rate of reaction \ce{N+ + H2}}

Reaction (1) has a small activation energy in the range 130-380~K
\citep{gerlich1993}, with a consensus now tending towards a value
below 200~K \citep[\eg\,][]{gerlich2008}. This reaction has been
studied experimentally by \cite{marquette1988} who considered pure
para-\hh\ and a 3:1 mixture of o- and \phh, refered to as normal \hh\
(\nhh). These authors fitted the two rates with normal and \phh\ as
$k_n = 4.16\tdix{-10}\,\exp[-41.9/T]\qquad \rm \cccs$ and $k_p =
8.35\tdix{-10}\,\exp[-168.5/T]\qquad \rm \cccs$.
The rate with p-\hh\ was used by LB91, who assumed that the reaction
with \ohh\ proceeds with no endothermicity, on the basis that the
170.51~K internal energy of the $J=1$ level of \ohh\ is used to
overcome the endothermicity. Accordingly, the rate of reaction
\reac{nhp} was written as $k_{\rm LB91} = 8.35\tdix{-10}
[x+(1-x)\exp(-168.5/T)]$, where $x=\opr/(1+\opr)$ is the fraction of
\ohh. Actually, the expressions for $k_n$ and $k_p$ allow to derive a
simple expression for $k_o$, the reaction rate of \ce{N+} with \ohh,
namely $k_n = 3/4\,k_o + 1/4\,k_p$ \citep{gerlich1989}. For \opr=3:1
and at temperatures less than $\approx40$~K, the rate is essentially
$k_o$, and $k_n$ is thus a good approximation of $k_o$ at the low
temperatures ($T<15$~K) of dark clouds. Alternatively, a
single-exponential fit to $k_o$ leads to (see Fig.~\ref{fig:ko})
\begin{eqnarray}
  \label{eq:ko}
  k_o &=& 4.2\tdix{-10}\,(T/300)^{-0.17}\,\exp[-44.5/T]\quad\rm \cccs
\end{eqnarray}
The rate for reaction \reac{nhp} with an \hh\ admixture of arbitrary
\opr\ is then obtained as
\begin{eqnarray}
  \label{eq:k1}
  k_1 = x\,k_o + (1-x)\,k_p
\end{eqnarray}
In Figure~\ref{fig:exp}, the rate $k_1$ from Eq.~\ref{eq:k1} is
compared to the experimental data of \cite{marquette1988} for the
reaction with n- and p- \ce{H2} down to 8~K and 45~K
respectively. \cite{gerlich1993} performed experiments with \nhh\ and
with \phh\ containing admixtures of \ohh\ (13\%, 8\%, and 0.8\%), at
temperatures down to 14 K. The agreement between our Eq.~\ref{eq:k1}
and the three sets of \phh\ data of \cite{gerlich1993} is excellent at
temperatures below 20~K, and to within a factor of 2 up to 100 K. This
suggests that Eq.~\ref{eq:k1} should be accurate to within a factor of
2-3 down to 10~K. On the other hand, this rate should not be employed
above $\sim 150$~K.

{We compare, in Fig.~\ref{fig:k1}, the rate given by Eq.~\ref{eq:k1}
  to $k_{\rm LB91}$ and to the rate given by the OSU database $k_{\rm
    OSU} = 8.35\tdix{-10}\,\exp[-85/T]$\,cm$^3$\,\pers.  The rate is
  computed for several values of \opr. As expected, it converges
  towards the n-\hh\ and p-\hh\ rates for large (\opr=3) and small
  (\opr=\dix{-4}) values respectively.  It is evident that LB91
  overestimated the rate of reaction (1) by several orders of
  magnitude at temperatures smaller than 20~K.}  In warmer gas
($T>20$~K), $k_1$ and $k_{\rm LB91}$ agree to within a factor of 10 or
less, whilst the OSU rate approaches the \opr=3 curve. The OSU rate
thus amounts to assume \opr\ ratios larger than 0.01, whereas LB91
provides acceptable rates only at temperatures larger than 20~K.


\subsection{{Abundance of nitrogen hydrides}}

\begin{table}[t]
  \centering
  \caption{Initial gas phase fractional abundances ($n(X)/\nh$) from \cite{flower2003} 
    with $\nh = n(\h) + 2n(\hh)$. For comparison, the fractional abundances from the
    low-metal model of \cite{wakelam2008} are also given. Numbers in parentheses are powers of 10.}
  \begin{tabular}{l r r}
    \hline
    \hline
    \rule[-1ex]{0mm}{4ex}
    Species & Flower 2003 & Wakelam 2008\\
    \hline
    \rule[0ex]{0mm}{3ex}%
    H$_2$  & 0.50     & 0.50    \\
    He     & 0.10     & 0.14    \\
    N      & 6.39(-5) & 2.14(-5)\\
    O      & 1.24(-4) & 1.76(-4)\\
    C$^+$  & 8.27(-5) & 7.30(-5)\\
    S$^+$  & 1.47(-5) & 8.00(-8)\\
    Si$^+$ & 8.00(-9) & 8.00(-9)\\
    Fe$^+$ & 3.00(-9) & 3.00(-9)\\
    \rule[-1.5ex]{0mm}{3.5ex}%
    Mg$^+$ & 7.00(-9) & 7.00(-9)\\
    \hline
  \end{tabular}
  \label{tab:abinit}
\end{table}

In this work, we wish to reproduce the ratios \ratio\ observed in the
cold envelope of \iras\ (Paper I). The fundamental hyperfine
transitions of NH, \ce{NH2} and several transitions of \ce{NH3} have
been detected in absorption. Absorption is interpreted as resulting
from the low temperature of the gas in the envelope seen against the
warmer continuum emitted by the dust closer to the protostar. The gas
density is low (\dix{4}-\dix{5}\,\ccc) as compared to the critical
density of the detected transitions ($> 10^7$\,\ccc) which ensures
that collisions do not govern the (de)excitation processes. The
$N=1-0$ transitions are therefore thermalized with the dust emission
temperature which, in the THz domain, correspond to excitation
temperatures close to a kinetic temperature of 10~K. The lines are
Gaussian and do not show signatures of strong dynamical effects such
as infall. Dynamical timescales are then expected to be large with
respect to the free-fall time. The observed lines therefore trace a
cold gas, moderately dense, free of dissociating photons, and where
the ionization is driven by cosmic rays.

The main point raised by Paper I is that the observed ratios
\ratio=5:1:300 could not be reproduced by chemical networks updated
regarding the rates of the dissociative recombination (DR) reactions
leading to NH, \ce{NH2}, and ammonia. Paper I considered three models,
with varying branching ratios for some DR reactions. Whilst the
NH$_2$:NH$_3$ abundance ratio could be reproduced in all three cases,
no model was able to produce\footnote{Hereafter, the fractional
  abundance of species X is noted \ab{X} and equals $n(\ce{X})/\nh$.}
$\ab{NH} > \ab{NH2}$. These models used the OSU rate for reaction
\reac{nhp} which was shown in Sec.~2.1 to depart from the measured
rate by several orders of magnitude in cold gas and \opr\ smaller than
0.01. The next Section explores the consequence of the new rate given
by Eq.~\ref{eq:k1} on the abundances of the nitrogen hydrides, in
typical dark cloud conditions.

We have performed chemical calculations in a gas at $T=10$~K, with
density $\nh=\dix{4}$\,\ccc. The gas is screened by 10~mag of visual
extinction, such that the ionization is primarily due to cosmic rays
at an adopted rate $\zeta = 1.3\tdix{-17}$\,\pers. Gas-phase
abundances are computed as a function of time by solving the chemical
network until steady-state is reached. Freeze-out of gas-phase species
onto dust grains are ignored in this work. In what follows, the quoted
abundances correspond to the steady-state. The initial fractional
elemental abundances ($n_{\rm X}/\nh$) are the gas phase abundances
taken from \citet[][ hereafter FPdF03]{flower2003} who considered the
depletion of metals in grain mantles, grain cores, and PAHs (see
Table~\ref{tab:abinit}). These abundances differ from those adopted in
Paper~I \citep{wakelam2008}. The rates for the dissociative
recombination reactions are those of the model 1 of Paper I. From the
above, the rate for the key reaction \reac{nhp} depends on the \opr\
which is not known, but which might strongly differ from the
thermodynamical equilibrium value \citep{maret2007, pagani2009,
  troscompt2009} which is $\approx \dix{-7}$ at 10~K. The steady-state
abundances of NH, \ce{NH2}, and \ce{NH3} have thus been computed for
various values of \opr\ in the range \dix{-7} to 3.

In this work, the DR of \ce{NH4+} has three output channels, \ce{NH3 +
  H}, \ce{NH2 + H2}, and \ce{NH2 + H + H}
\citep{ojekull2004}. However, another channel may be \ce{NH + H + H2}
\citep{adams1991}, but to our knowledge, no branching ratio is
available in the litterature. Values up to 10\% may be considered in a
future work.

\begin{figure}
  \centering
  \includegraphics[height=\hsize,angle=-90]{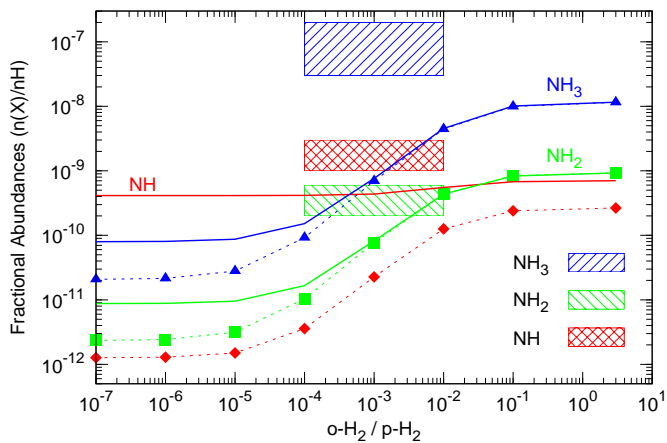}
  \caption{Steady-state abundances (with respect to H nuclei) of
    nitrogen hydrides as a function of \opr, in a 10~K gas with
    \nh=\dix{4}\,\ccc, and $\zeta=1.3\tdix{-17}$~\pers. The rate of
    reaction (1) is given by Eq.~\ref{eq:k1}. Two branching ratios for
    the channel \ce{N2H+ + e- -> NH + N} are considered: 0\% (dashed
    lines) and 10\% (full lines). The observed abundances from Paper~I
    are also indicated (filled rectangles).}
  \label{fig:xab}
\end{figure}

We first consider the case where the DR of \ce{N2H+} produces only
\ce{N2 + H} \citep{adams2009}. The resulting abundances are the dashed
lines in Fig.~\ref{fig:xab}. It appears that the \opr\ controls the
abundances of \ce{NH}, \ce{NH2}, and \ce{NH3} but the ratios \ri\ and
\rii\ are insensitive to \opr. In a second series of calculations, the
\ce{NH + N} channel is given a 10\% branching ratio for the DR of
\ce{N2H+}. In this case, the abundance of NH is nearly independent of
\opr, while the abundances of \ce{NH2} and \ce{NH3} remain unaffected
(full lines in Fig.~\ref{fig:xab}. As a consequence, it is found that
for $\opr < 0.03$, $\ri > 1$. In Paper I though, opening this channel
did not solve the NH:\ce{NH2} problem, which contrasts with the above
result. This may be understood as follows.  When the \ce{NH + H}
channel of the dissociative recombination of \ce{N2H+} is opened with
a 10\% braching ratio, the corresponding rate is $\approx
5.5\tdix{-8}$\,\cccs, several orders of magnitude larger than $k_1$
(see Fig.~\ref{fig:k1}). Hence this channel, which is insensitive to
\opr, dominates the formation of \ce{NH} over \ce{NH2+ + e-}. In
contrast, the other hydrides \ce{NH2} and \ce{NH3} are daughter
molecules of \ce{NH+} which is formed from \ce{N+ + H2} whose rate
does explicitely depend on \opr\ in the present work. Hence the
abundances of \ce{NH2} and \ce{NH3} do also depend on \opr. On the
other hand, at the temperature of 10~K and $\opr < \dix{-3}$, the
revised rate $k_1$ (Eq.~\ref{eq:k1}) is at least an order of magnitude
smaller than the OSU rate used in Paper~I. For larger \opr, the rate
is similar or even larger up to a factor of 10.  These two effects,
namely the drop of $k_1$ for low \opr, and the \opr-independant
formation of NH, makes it possible to produce a \ri\ ratio with values
greater than unity. In the following we keep the \ce{NH + N} channel
opened to 10\% and explore some consequences of this result.

\section{Discussion}

\subsection{The \hh\ \opr\ ratio in dark clouds}

Computing the abundance ratios \ri\ and \rii\ for different \opr\
ratios leads to the left panel of Fig.~\ref{fig:ratios}. The initial
fractional abundances are kept fixed, while the rate of reaction
\reac{nhp} is calculated for values of \opr\ from \dix{-7} (close to
the Boltzmann value at 10~K) to 3. The calculated ratios are compared
with the observational constraints. As expected, the \rii\ ratio is
insensitive to \opr\ variations. It is equal to 10, not consistent
with the observed value of 300. On the contrary, the \ri\ ratio shows
two regimes: for $\opr < \dix{-5}$, $\ri \approx 50$, and for $\opr >
0.1$, $\ri \approx 0.07$. The observed ratio of 5 is intermediate, and
indeed well constrains the \opr\ to \dix{-3}. From
Fig.~\ref{fig:ratios}, we conclude that there is a range of \opr\ for
which $\ab{NH}>\ab{NH2}$ and $\ab{NH}<\ab{NH3}$ simultaneously. In our
case, this range is $\opr=5\tdix{-4}-0.02$.  The \opr\ ratio appears
as a control parameter for the \ri\ ratio, and allows for the first
time to produce more NH than \ce{NH2}. Conversely, the measure of \ri\
tightly constrains the \opr\ ratio. We note that the fractional
abundance of \ce{NH3} of 1.9\tdix{-9}, measured by \cite{crapsi2007}
towards the starless core L~1544, corresponds to an
$\opr=\dix{-3}-\dix{-2}$. This ratio is a factor of 10 larger than the
corresponding ratio from the model of LB91. It is thus found that
reaction \reac{nhp} regulates the formation of \ce{NH2} and \ce{NH3}
whilst the formation of NH is controlled by the dissociative
recombination of \ce{N2H+} provided the branching ratio towards NH is
10\%. The value of this branching ratio is uncertain but is likely
non-zero \citep{adams2009}.

\subsection{Dependence on the initial abundances}

The \rii\ ratio is $\approx 10$ in the above models, a factor 30 below
the observed values towards \iras. However, models computed in Paper~I
with the initial abundances of \cite{wakelam2008} lead to a ratio
close to the observed value, hence suggesting that the initial
elemental abundances influence this ratio. The gas phase elemental
abundances of C, N, and O in dark clouds are poorly known because the
amount of these elements incorporated into the dust (core or mantles)
is loosely constrained. The elemental abundance of oxygen in the gas
phase is not known accurately, and variations by an order of magnitude
are fully conceivable \citep{jenkins2009}, whilst that of carbon is
better known.  Accordingly, we have considered variations of the
initial elemental abundances of oxygen, to vary the ratio C:O,
encompassing the value of 0.41 from \cite{wakelam2008}.

\begin{figure*}[t]
  \centering
  \includegraphics[height=0.5\hsize,angle=-90]{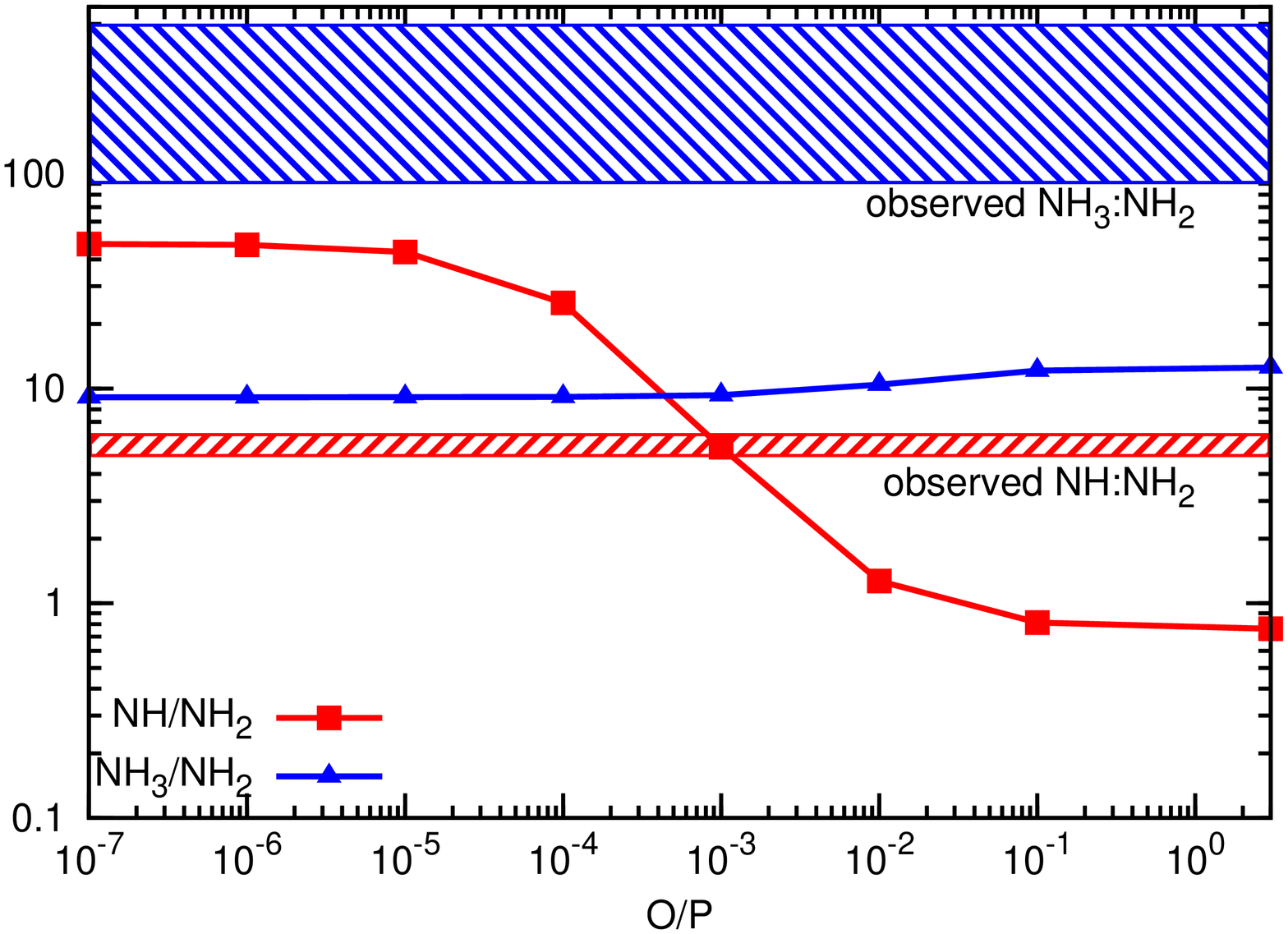}\hfill%
  \includegraphics[height=0.5\hsize,angle=-90]{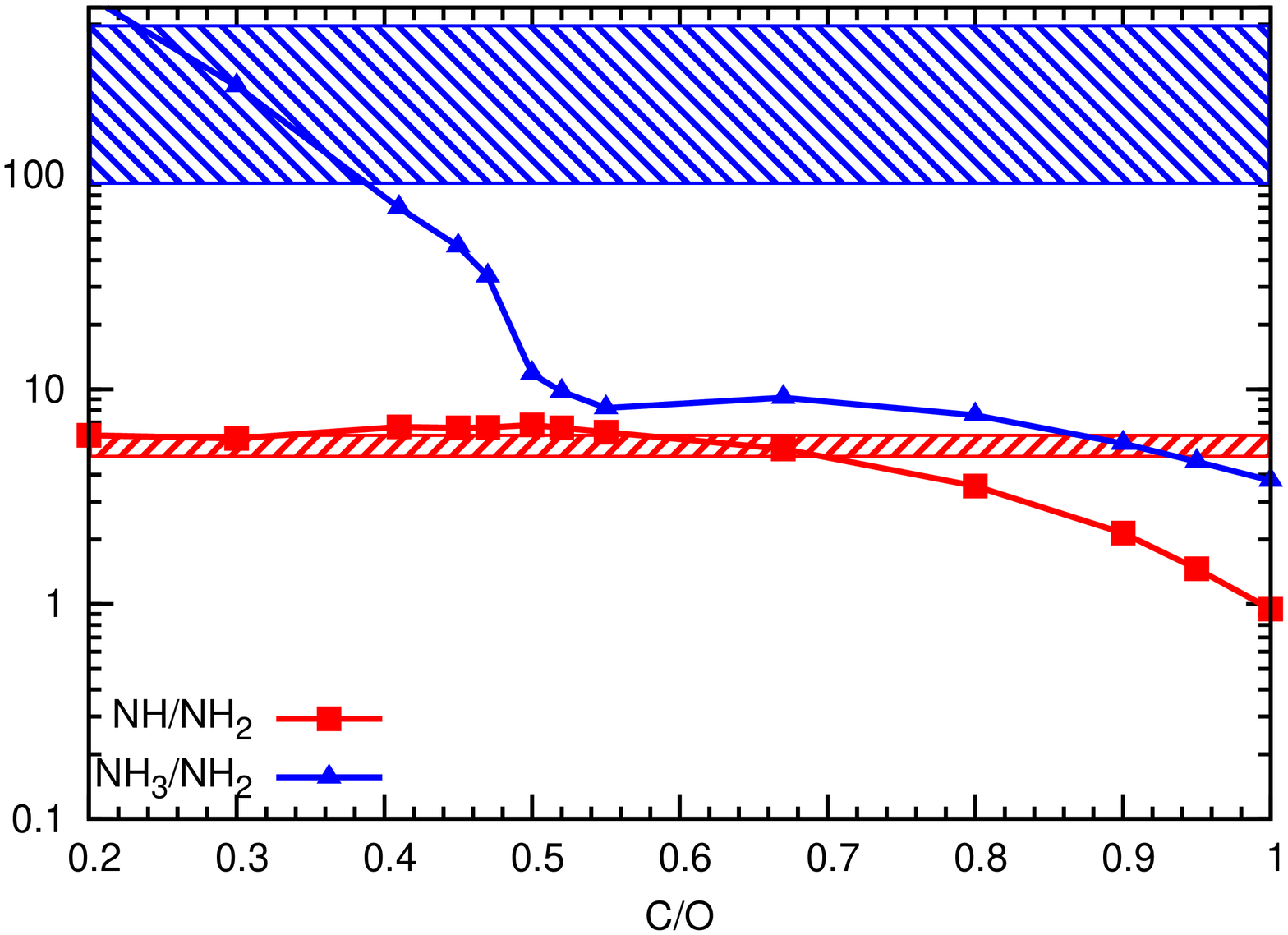}
  \caption{Abundance ratios calculated when a 10\% branching ratio is
    considered for the reaction \ce{N2H+ + e- -> NH + H}. The hashed
    bands show the observed ratios with their $1\sigma$
    uncertainties. \textit{Left panel}: varying \opr, with the initial
    abundances of \cite{flower2003} (\ab{O}=1.24(-4) and
    \ab{C}=8.27(-5).). \textit{Right panel}: varying \ab{O} while
    keeping \ab{C} constant, at constant \opr=\dix{-3}. The low-metal
    abundances of \cite{wakelam2008} are \ab{O}=1.76(-4) and
    \ab{C}=7.30(-5), or C:O=0.41.}
  \label{fig:ratios}
\end{figure*}

The resulting abundance ratios, computed for an \opr\ ratio of
\dix{-3}, are shown in Fig.~\ref{fig:ratios} (right panel). As C:O
decreases below the 0.66 value of FPdF03, the ratio \ce{NH2}:\ce{NH3}
increases. In the process, \ce{NH}:\ce{NH2} remains constant. When C:O
is now increased above 0.66, both ratios decrease, by at most a factor
2 to 5. It is thus apparent that, when \ab{O} is increased, the C:O
ratio controls the \ce{NH2}:\ce{NH3} ratio. On chemical grounds, the
C:O ratio is expected to influence the abundance of \ce{NH2}, for
which the main destruction routes involve oxygen to form NO, NH, and
HNO. Similarly, NH is mostly removed by reaction with oxygen to form
principally NO. However, when oxygen is significantly depleted from
the gas phase (C:O $> 0.5$), another destruction route of NH,
involving sulfur, becomes important. As a result, the \ri\ ratio is
mostly insensitive to C:O, until C:O$>0.5$ when it starts to decrease
by small factors.



The situation is different for \ce{NH3} which is destroyed by charge
transfer reactions with \ce{H+} and \ce{S+} and by proton exchange
reactions with \ce{H3+} and \ce{HCO+}, forming notably \ce{NH4+},
which dissociates back into \ce{NH3}, and \ce{NH3+}. A fraction of
\ce{NH3+} will lead to \ce{NH2} and \ce{NH} which are the true
destruction channels of ammonia. The abundance of \ce{NH3} is thus
only marginally affected by the change in C:O, unless the ionization
fraction is modified, which is the case for low C:O ratios. An
increase of the oxygen abundance (\ie\ a decrease of C:O) is
accompanied by a deacrease of \ab{e-}, or equivalently \ab{S+} -- the
dominant ion -- which makes the abundance of \ce{NH3} to
increase. Consequently, as C:O decreases below 0.66, \ri\ keeps
constant and \rii\ increases by more than one order of
magnitude. Whereas when C:O increases, \ri\ decreases by less than a
factor 10, whilst \rii\ is approximately constant.

\section{Conclusions}

Using an updated rate for reaction \reac{nhp} with an explicit
dependence on the \opr\ ratio, we have shown that the \opr\ of \hh\
controls the ratio \ri\ in dark clouds without affecting the \rii\
ratio. A value of \opr\ close to \dix{-3} leads to $\ri >1$ and
$\rii<1$ as observed. Interestingly, this value of \opr\ is close to
the predictions of \citet[][their Fig.~1]{flower2006a} under similar
conditions. In addition, measuring the \ri\ ratio may be a new method
to constrain the \opr\ ratio of \hh\ in dark clouds. We have also
shown that decreasing the C:O ratio by increasing \ab{O} controls the
\ce{NH2}:\ce{NH3} ratio. Finally, acceptable parameters are found that
lead to abundance ratios in agreement with the observations towards
\iras, namely $\opr\approx\dix{-3}$ and C:O$\le0.4$.  It is to be
noted, however, that in this range of parameters, the absolute
abundances predicted by the models are a factor of 10 below those
derived in Paper~I.


In this work, the influence of \opr\ on the ratios \ratio\ has been
explored only through the rate of reaction \reac{nhp}.  A better study
would be to self-consistently compute the \opr\ ratio from a model of
conversion of \ohh\ into \phh, as was first considered by LB91, and
more recently by \eg\ \cite{pagani2009}. Including proton exchange
reactions, in the gas phase, between \hh\ and \ce{H+}
\citep{honvault2011}, \ce{H2+} \citep{crabtree2011}, and \ce{H3+}
\citep{hugo2009}, would reprocess the 3:1 mixture of \hh\ formed on
grains to a different \opr. An even more self-consistent approach
would be that of \cite{flower2006a} who separate reactions with the
ortho- and para- forms of all concerned N-bearing molecules.  Another
obvious continuation would be to explore these findings in different
physical conditions {(\av, cosmic-ray ionization rate, \etc)}.

{Branching ratios of dissociative recombination reactions are
  crucial to the formation of nitrogen hydrides. In this work, the
  dissociative recombination of \ce{NH4+} has three output channels
  \ce{NH3 + H}, \ce{NH2 + H2}, and \ce{NH2 + H + H}
  \citep{ojekull2004}. However, another channel may be \ce{NH + H +
    H2} \citep{adams1991}, but to our knowledge, no branching ratio is
  available in the litterature. Values up to 10\% may be considered in
  a future work. Critical too are the values of rates at low
  temperatures, especially for reaction (1) with \phh\ for which
  measurements at temperatures lower than 14~K are not available.}

\acknowledgement{We thank Pr. D.~Gerlich for stimulating discussions
  and an anonymous referee for useful comments which improved the
  manuscript. We acknowledge financial support from the CNRS national
  program ``Physique et Chimie du Milieu Interstellaire''.}

\bibliographystyle{aa}
\bibliography{general,cores,chemistry,phb,technic,disks}

\Online

\appendix

\section{Reaction rate}

\begin{figure}
  \centering
  \includegraphics[width=\hsize]{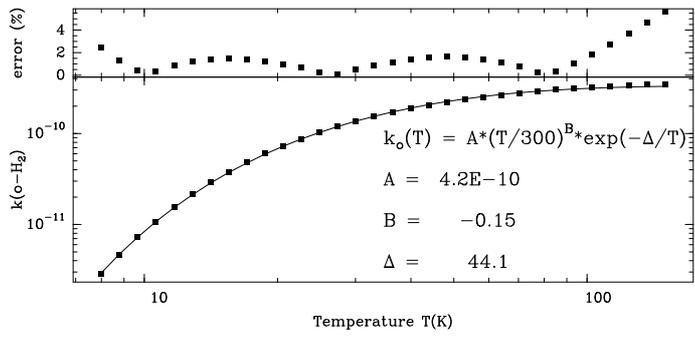}
  \caption{Rate for reaction \reac{nhp} with \ohh\ and fit result (see
    Eq.~\ref{eq:ko}). Bottom panel: filled squares are the values
    derived from \cite{marquette1988}, and the fitted curve (full
    line). Top panel: relative error.}
  \label{fig:ko}
\end{figure}

For the purpose of implementing the rate of the reaction \reac{nhp}
with ortho-\hh\ in chemical networks, Figure~\ref{fig:ko} shows the
result of a single-exponential fit. The filled squares are based on
the fitted reaction rates from \cite{marquette1988}. A weighting in
the form $1/(T-10)^2$ was adopted, which is accurate to better than
6\% for $T=8-150$~K.

\end{document}